\documentclass[%
 preprint,%
 amssymb, amsmath,%
 prl,cha,superscriptaddress,%
]{revtex4-1}
\usepackage{graphicx}
\usepackage{comment}
\usepackage[capbesideposition=right]{floatrow}
\usepackage{sidecap}
\usepackage{nicefrac}
\usepackage{docs}%
\usepackage{bm}%
\usepackage[colorlinks=true,linkcolor=blue]{hyperref}%
\expandafter\ifx\csname package@font\endcsname\relax\else
\expandafter\expandafter
\expandafter\usepackage
\expandafter\expandafter
\expandafter{\csname package@font\endcsname}%
\fi
\begin{document}

\title{Angular dependence of the domain wall depinning field in sensors with segmented corners.}

\author{Daniel Heinze}%

\affiliation{Institut f\"ur Physik, Johannes Gutenberg-Universit\"at Mainz, Staudinger Weg 7, 55128 Mainz, Germany. Contact: klaeui@uni-mainz.de}%

\author{Benjamin Borie}%

\affiliation{Institut f\"ur Physik, Johannes Gutenberg-Universit\"at Mainz, Staudinger Weg 7, 55128 Mainz, Germany. Contact: klaeui@uni-mainz.de}%
\affiliation{Sensitec GmbH, Hechtsheimer Str. 2, Mainz D-55131, Germany}%

\author{J\"urgen Wahrhusen}%
\affiliation{Sensitec GmbH, Hechtsheimer Str. 2, Mainz D-55131, Germany}%

\author{Hubert Grimm}%
\affiliation{Sensitec GmbH, Hechtsheimer Str. 2, Mainz D-55131, Germany}%

\author{Mathias Kl\"aui}%
\affiliation{Institut f\"ur Physik, Johannes Gutenberg-Universit\"at Mainz, Staudinger Weg 7, 55128 Mainz, Germany. Contact: klaeui@uni-mainz.de}%

\begin{abstract}

Rotating domain wall based sensors that have recently been developed are based on a segmented looping geometry. In order to determine the crucial pinning of domain walls in this special geometry, we investigate the depinning under different angles of an applied magnetic field and obtain the angular dependence of the depinning field of the domain walls. Due to the geometry, the depinning field not only exhibits a 180$^\circ$-periodicity but a more complex dependence on the angle. The depinning field depends on two different angles associated with the initial state and the segmented geometry of the corner. We find that depending on the angle of the applied field two different switching processes occur and we can reproduce the angular dependence using a simple model calculation.

\end{abstract}

\maketitle

\section{Introduction}

Magnetic domain walls are a of great interest for the sensor industry for their ability to be stable at room temperature and to be easily moved by magnetic fields [\cite{Kla08},\cite{Die04}]. Currently, only one sensor based on domain walls is produced and sold [\cite{Die04}, \cite{Novo}]. One of the difficulties that needs to be overcome to use domain walls in more devices is the complex dynamics of the domain wall behaviour in structures with different geometries [\cite{Bis13},\cite{Ric16}]. The sensor devices are characterized by a magnetic operating window set by the nucleation field (upper bound) and the depinning or propagation field (lower bound) in which the sensors function. The propagation and depinning of domain walls is affected by various internal and external effects. The edge roughness [\cite{Chi10},\cite{Mar07}], the crystallite size \cite{Voto16} and the variations of the topography can affect the domain wall. In addition, the Walker Breakdown \cite{Walk75} changes the spin structure of the domain wall rendering the determination of the pinning challenging. In the case of straight wires, kinks and rings, the motion and depinning of domain walls was intensively investigated [\cite{Mat12},\cite{Bed07},\cite{Cor15}]. However, polygons and segmented shapes can bring different results as compared with smooth geometrical variations and these geometries are used in the sensor devices. In earlier studies, it was found that the depinning field generally exhibits a pronounced angular dependence [\cite{Bed07},\cite{Cor15}]. In \cite{Bed07}, it was described that the switching field is lowest when the field is tangential to the circumference of the structure at the domain wall position. The measurement was performed in that case on circular rings without sharp variations in the path of the domain wall. However, the influence of sharp corners in segmented loops will impact the pinning and depinning processes and needs to be determined to ascertain the most reliable structures for domain wall displacements.

In this paper, we report the Kerr-microscopy measurement of domain wall depinning from corners in a segmented looping structure under different angles of an applied magnetic field. We show that the depinning field is a function of the initial position and the segmented geometry and explain it based on an analysis of the domain wall motion for different angles.

The samples were produced on a wafer of Si/AlO$_x$ and deposited in a magnetron sputtering tool. The sputtered magnetic stack consists of 1 nm of CoFe, 29 nm of NiFe and a capping layer of 4 nm of Ta. The shape of the device was then obtained through a photolithography step and an ion milling using Ar ions. A width of 350 nm is obtained with an edge roughness of typically 10 nm. A sketch of the geometry is shown in \autoref{fig:samplesketch}.

\vspace{0,5cm}

\section{Measurement of the angular dependent depinning field}

\vspace{0,5cm}

\begin{figure}[H]
\centering
\includegraphics[scale=0.78]{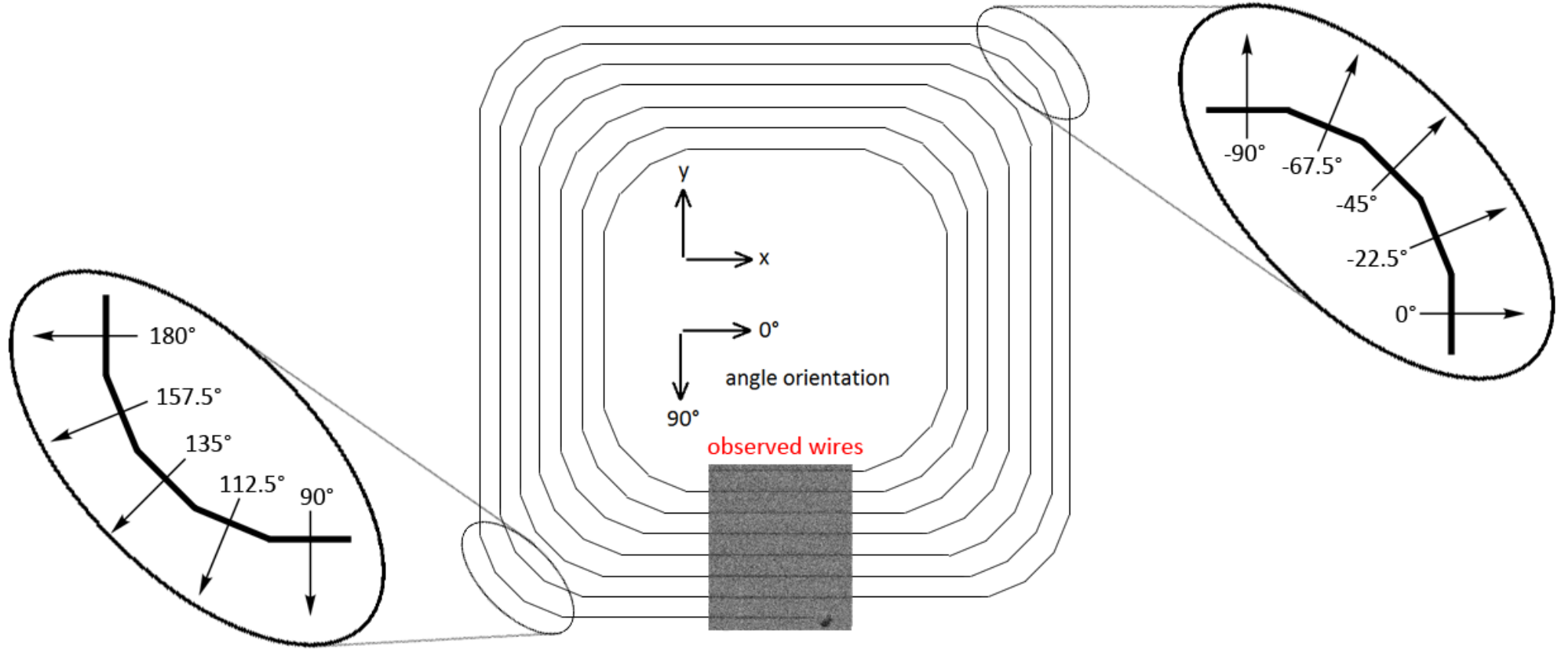}
\caption{{\small Sketch of a sensor with zoomed in bottom left and top right corners. The angle orientation and the definition of the x- and y-axis to simplify explanations is also given. A Kerr microscopy imaged shows the observed wires with the magnetic contrast visible.}}
\label{fig:samplesketch}
\end{figure}

The measurements were performed on magnetic domain wall based rotation sensors, which have a square-shaped geometry with segmented corners. They consist of 16 loops in total starting with a nucleation pad connected to the outer wire (\autoref{fig:samplesketch}). The technique used to image is a magneto-optical Kerr effect microscope in the longitudinal mode \cite{Scha07}. The orientation of the sensor under the microscope is kept fixed for the experiment as shown in \autoref{fig:samplesketch}. A magnetic field is applied and the angle between the field and the imaged stripes is defined as 0$^\circ$ points in the x direction and at 90$^\circ$ is the -y direction.

\vspace{0,5cm}

\vspace{0,5cm}

To carry out the depinning experiment, domain walls are nucleated in the structure and positioned in the segmented corners. This is done by the application of a strong magnetic field along an angle of 135$^\circ$ (\autoref{fig:samplesketch}). When the magnetic field is reduced to zero, the magnetization relaxes to a state with domain walls in the top right and bottom left corner that are shown in \autoref{fig:samplesketch}.

\begin{figure}[H]
\centering
\includegraphics[scale=0.78]{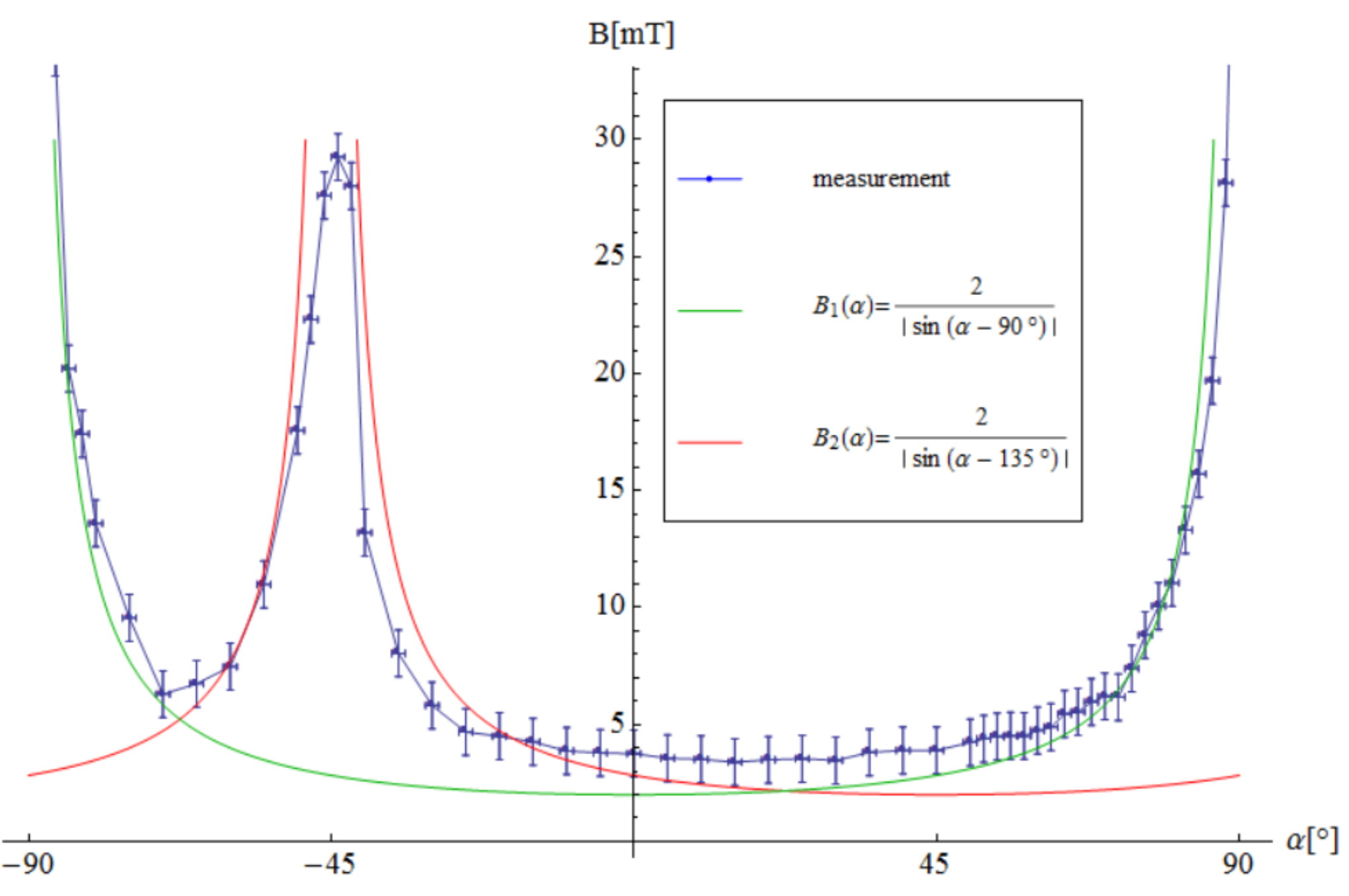}
\caption{{\small Measurement and fit of the angular dependency of the switching field.}}
\label{fig:depinningtotalfit}
\end{figure}

After the initialization, the angle of the applied field with the imaged stripes was set to a constant value between 90$^\circ$ and -90$^\circ$ and the field was increased until a contrast change could be observed in the observed wires. The angle was then varied by 2$^\circ$ and the operation was repeated. For each angle this procedure was repeated ten times, so at each angle there was an average of 80 values due to the eight wires observed by the Kerr microscopy imaging (see \autoref{fig:samplesketch}). For our purposes the mean of the depinning field was taken from the observed eight wires for each angle. The result of the measurement of the field at which the wires reverse as a function of the applied field direction can be seen in \autoref{fig:depinningtotalfit}.

\vspace{0,5cm}

\vspace{0,5cm}

To simplify the analysis, only the abrupt reversal in the observed wires was detected and the field values for this event are plotted. 
First, we see that the depinning field diverges at $\pm$90$^\circ$, this can be explained by the Zeeman energy and is due to the probed field of view of the sensor.

\vspace{0,5cm}

\begin{equation}
	E_{\text{Zeeman}}=-\mu_0\int_V{\vec{M}\cdot\vec{H}_{\text{ext}}\,\text{d}V}
\label{eq:zeeman}
\end{equation}

\vspace{0,5cm}

In \autoref{eq:zeeman}, $\vec{M}$ is the magnetization and $\vec{H}_{\text{ext}}$ is the applied external magnetic field. For the magnetization of the wires being aligned along the wire axis  due to the shape anisotropy, the applied field is orthogonal to the magnetization. Therefore, neither depinning nor nucleation of the domain walls can be imaged in the observed wires for these field directions. To explain the angular dependence, we divide the graph into different sections and in particular the angle range between 90$^\circ$ and -90$^\circ$ is divided in three different parts based on the movement of the different domain walls and the resulting switching processes occurring for the different field angles.

\begin{figure}[H]
\centering
\includegraphics[scale=0.78]{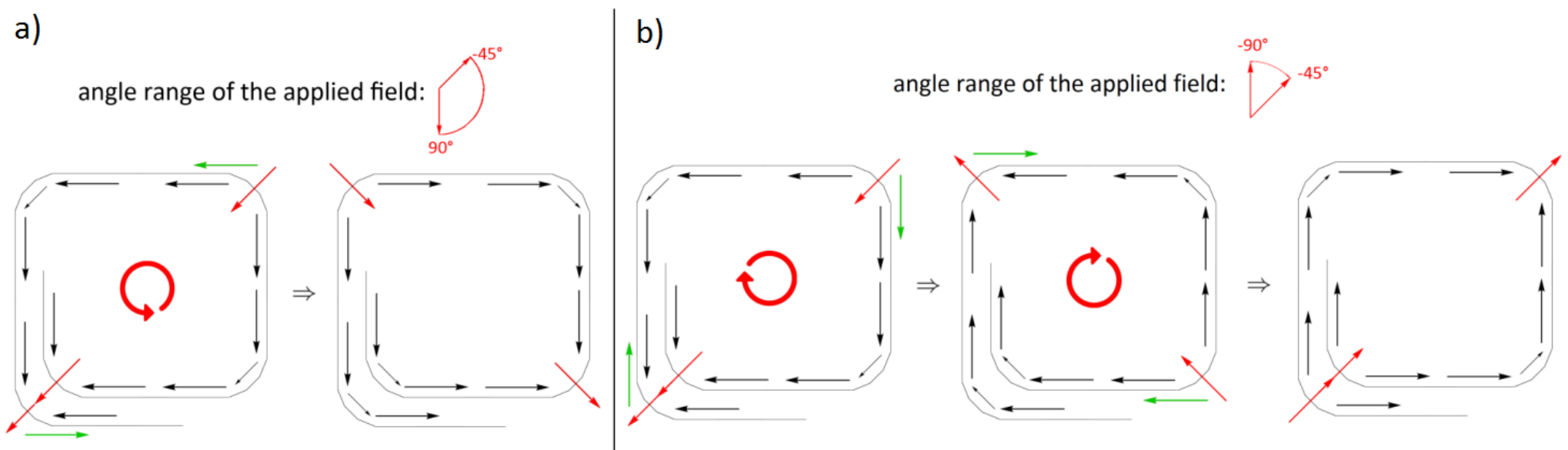}%
\caption{{\small a) Sketch of the domain wall propagation in the range of 90$^\circ$ to -45$^\circ$. b) Sketch of the two-step domain wall propagation process in the range of -45$^\circ$ to -90$^\circ$. The domain walls are indicated by red arrows and their movement directions by green arrows. The black arrows indicate the magnetization directions in the looped wires. The opposite displacement direction for a) and b) is indicated as well in thick red.}}
\label{fig:depinning}
\end{figure}

(I) In the first range between 90$^\circ$ and 0$^\circ$, the -y-component is in the same direction as the vertical parts of the loops. The domain walls located in the corners will move horizontally as schematically shown in \autoref{fig:depinning}a where the displacement direction of the walls is shown by green arrows.

(II) In the range between 0$^\circ$ and -45$^\circ$,  the x-component of the applied magnetic field is stronger than the y-component and both are in direction opposite to the current magnetization configuration at the domain wall position.  Therefore, the domain walls at the bottom left corner move first to the right  and the ones in top right to the left thus changing the magnetization direction in the observed wires (\autoref{fig:depinning}a). The motion is thus anti-clockwise. In a second step, the domain wall in the top wire moves down and the one in the bottom wire moves up. This part is not shown schematically as it is not imaged because the magnetization in the observed wires (\autoref{fig:samplesketch}) is not affected.

(III) Finally, between -45$^\circ$ and -90$^\circ$, the domain walls also move in a two-step process. First, they move in the vertical direction because the y-component of the field is stronger than the x-component (first step in \autoref{fig:depinning}b). Then, if the x-component of the applied field is sufficiently high, the domain walls can also depin from the new position in the corners in a second step (also shown in \autoref{fig:depinning}b) and move horizontally and therefore a contrast change in the observed wires can be detected. The sense of rotation of the wall movement for this angle range is clockwise.

(IV) An interesting feature appears at  -45$^\circ$. At this angle, the applied magnetic field is  exactly anti-parallel to the initial position of the domain walls. Therefore, at -45$^\circ$, the Zeeman force pushing the domain wall horizontally is equal to the one pushing it vertically thus the total acting torques lead to the domain wall being in a metastable equilibrium position. But the measured field values for the switching do not completely diverge. There are two possible explanations for this behaviour: The nucleation of new domain walls or the depinning due to inevitable asymmetries in the geometry and domain wall structure. The detected field value in our system is 29 mT, which is also the field needed to nucleate new domains starting from a spin structure without domain walls so that we identify nucleation as the process.

\vspace{0,5cm}

\vspace{0,5cm}

Furthermore we see that the local minimum in the depinning field at -67.5$^\circ$ is not as low as at -22.5$^\circ$. This asymmetry can be explained by the fact that the depinning field from the -45$^\circ$ angle goes down while the field is rotated to -90$^\circ$. However, as the field approaches -90$^\circ$, the detection of the depinning of domain walls in the observed wires becomes increasingly difficult. 

Generally the depinning field depends on the component acting in the direction of the displacement (sine dependence) as shown in \cite{Bed07}. As a result of the geometry, the depinning field depends here on two angles: Since the sensors are square-shaped, the angular dependence has a 180$^\circ$-periodicity, and due to the position of the observed wires, there is a divergence at $\pm$90$^\circ$ (\autoref{eq:b1}). Furthermore due to the positioning of the domain walls in the initial state (IV) there is the divergence at 135$^\circ$ and -45$^\circ$, which results in another depinning field, depending on this angle (\autoref{eq:b2}):

\vspace{0,5cm}

{\parindent0pt
	\begin{minipage}{\linewidth}
		\begin{minipage}{.5\linewidth}
		\centering
			\begin{equation}
				B_1(\alpha)=\left|\frac{2 mT}{\sin(\alpha-90^\circ)}\right|
			\label{eq:b1}
			\end{equation}
		\end{minipage}
		\hfill
		\begin{minipage}{.5\linewidth}
			\centering
			\begin{equation}
				B_2(\alpha)=\left|\frac{2 mT}{\sin(\alpha-135^\circ)}\right|
			\label{eq:b2}
			\end{equation}
		\end{minipage}
	\end{minipage}
}

\vspace{0,5cm}

We see that the depinning field is always the maximum of the two fields and this reproduces the experiment well (\autoref{fig:depinningtotalfit}). This also shows that in our geometry the depinning is different from previous studies with simpler geometries \cite{Bed07}. There, the minimal depinning field was with the applied field tangential to the wire, where the domain wall is located. In our case that would be at an angle of 45$^\circ$. This deviation is a result of the different shape of the sensor. In the circular wire (ring), all directions are equal. In contrast, with the square-shaped sensor and the segmented corner, there are selected directions that dominate the pinning governed by the directions of the segments (see \autoref{fig:samplesketch}), which results in the lowest depinning field at 22.5$^\circ$.

\vspace{0,5cm}

\section{Conclusion and prospects}

\vspace{0,5cm}

Summarizing, the angular dependence of the depinning and propagation of magnetic domain walls is highly geometry related. In particular the observed wires (\autoref{fig:samplesketch}) reverse by the domain walls arriving for different angle-ranges from different directions. Between 90$^\circ$ and -45$^\circ$ the domain walls propagate directly through the observed wires, due to the initial domain wall position ((I),(II)). In the range from -45$^\circ$ to -90$^\circ$ there is a two-step process necessary to change the magnetization in the observed wires (III), which leads to a local depinning field minimum at -67.5$^\circ$. With our square-shaped sensors and the initializing field at 135$^\circ$, at the two angles $\pm$90$^\circ$ and -45$^\circ$ no reversal by depinning and propagation of domain walls is observed. In contrast to $\pm$90$^\circ$, at -45$^\circ$ the nucleation of domain walls in the wires governs the switching of the magnetization. The superposition of the depinning fields for these angles yield a minimum at 22.5$^\circ$, which is not tangential to the wires, where the domain walls are initialized, showing a clear difference to the behaviour of circular rings, which can be ascribed to the segmented corners.

Further investigations have to reveal, if the peak at  -45$^\circ$ is only related to the initialization. This can be achieved by initializing the sensor for example at 112.5$^\circ$ instead of 135$^\circ$ (\autoref{fig:samplesketch}). If the peak occurs at -67.5$^\circ$, the theory of the correlation between initial state and the peak will be confirmed for this sensor geometry. The understanding of the depinning mechanism in segmented loops will improve the magnetic operating window. Indeed, an engineering of the number of segments is likely to lead to a lower depinning field and a larger operating window.

The authors kindly thank the company Sensitec GmbH for providing the samples, in particular J. Paul for fruitful discussions. The work and results reported in this publication were obtained with research funding from the European Community under the Seventh Framework Programme - The people Programme, Multi-ITN “WALL” Contract Number Grant agreement no.: 608031, the ERC (MultiRev ERC-2014-PoC (665672)) and the DFG (SFB TRR 173 Spin+X).

\vspace{0,5cm}
\bibliographystyle{apsrev4-1}
\bibliography{PaperHeinzeBorie}

%merlin.mbs apsrev4-1.bst 2010-07-25 4.21a (PWD, AO, DPC) hacked
%Control: key (0)
%Control: author (72) initials jnrlst
%Control: editor formatted (1) identically to author
%Control: production of article title (-1) disabled
%Control: page (0) single
%Control: year (1) truncated
%Control: production of eprint (0) enabled
\begin{thebibliography}{13}%
\makeatletter
\providecommand \@ifxundefined [1]{%
 \@ifx{#1\undefined}
}%
\providecommand \@ifnum [1]{%
 \ifnum #1\expandafter \@firstoftwo
 \else \expandafter \@secondoftwo
 \fi
}%
\providecommand \@ifx [1]{%
 \ifx #1\expandafter \@firstoftwo
 \else \expandafter \@secondoftwo
 \fi
}%
\providecommand \natexlab [1]{#1}%
\providecommand \enquote  [1]{``#1''}%
\providecommand \bibnamefont  [1]{#1}%
\providecommand \bibfnamefont [1]{#1}%
\providecommand \citenamefont [1]{#1}%
\providecommand \href@noop [0]{\@secondoftwo}%
\providecommand \href [0]{\begingroup \@sanitize@url \@href}%
\providecommand \@href[1]{\@@startlink{#1}\@@href}%
\providecommand \@@href[1]{\endgroup#1\@@endlink}%
\providecommand \@sanitize@url [0]{\catcode `\\12\catcode `\$12\catcode
  `\&12\catcode `\#12\catcode `\^12\catcode `\_12\catcode `\%12\relax}%
\providecommand \@@startlink[1]{}%
\providecommand \@@endlink[0]{}%
\providecommand \url  [0]{\begingroup\@sanitize@url \@url }%
\providecommand \@url [1]{\endgroup\@href {#1}{\urlprefix }}%
\providecommand \urlprefix  [0]{URL }%
\providecommand \Eprint [0]{\href }%
\providecommand \doibase [0]{http://dx.doi.org/}%
\providecommand \selectlanguage [0]{\@gobble}%
\providecommand \bibinfo  [0]{\@secondoftwo}%
\providecommand \bibfield  [0]{\@secondoftwo}%
\providecommand \translation [1]{[#1]}%
\providecommand \BibitemOpen [0]{}%
\providecommand \bibitemStop [0]{}%
\providecommand \bibitemNoStop [0]{.\EOS\space}%
\providecommand \EOS [0]{\spacefactor3000\relax}%
\providecommand \BibitemShut  [1]{\csname bibitem#1\endcsname}%
\let\auto@bib@innerbib\@empty
%</preamble>
\bibitem [{\citenamefont {Kl\"aui}(2008)}]{Kla08}%
  \BibitemOpen
  \bibfield  {author} {\bibinfo {author} {\bibfnamefont {M.}~\bibnamefont
  {Kl\"aui}},\ }\href@noop {} {\bibfield  {journal} {\bibinfo  {journal}
  {\textit{J. Phys.: Condens. Matter}}\ }\textbf {\bibinfo {volume} {20}},\
  \bibinfo {pages} {313001} (\bibinfo {year} {2008})}\BibitemShut {NoStop}%
\bibitem [{\citenamefont {Diegel}\ \emph {et~al.}(2004)\citenamefont {Diegel},
  \citenamefont {Mattheis},\ and\ \citenamefont {Halder}}]{Die04}%
  \BibitemOpen
  \bibfield  {author} {\bibinfo {author} {\bibfnamefont {M.}~\bibnamefont
  {Diegel}}, \bibinfo {author} {\bibfnamefont {R.}~\bibnamefont {Mattheis}}, \
  and\ \bibinfo {author} {\bibfnamefont {E.}~\bibnamefont {Halder}},\
  }\href@noop {} {\bibfield  {journal} {\bibinfo  {journal} {\textit{IEEE
  Trans. Magn.}}\ }\textbf {\bibinfo {volume} {40}},\ \bibinfo {pages} {2655}
  (\bibinfo {year} {2004})}\BibitemShut {NoStop}%
\bibitem [{Nov()}]{Novo}%
  \BibitemOpen
  \href@noop {} {\enquote {\bibinfo {title} {Novotechnik : Rsm-2800
  multiturn},}\ }\bibinfo {howpublished}
  {\url{http://www.novotechnik.de/produkte/winkelsensoren/}},\ \bibinfo {note}
  {accessed: 2016-07-17}\BibitemShut {NoStop}%
\bibitem [{\citenamefont {Bisig}\ \emph {et~al.}(2013)\citenamefont {Bisig},
  \citenamefont {St{\"a}rk}, \citenamefont {Mawass}, \citenamefont {Moutafis},
  \citenamefont {Rhensius}, \citenamefont {Heidler}, \citenamefont
  {B{\"u}ttner}, \citenamefont {Noske}, \citenamefont {Weigand}, \citenamefont
  {Eisebitt} \emph {et~al.}}]{Bis13}%
  \BibitemOpen
  \bibfield  {author} {\bibinfo {author} {\bibfnamefont {A.}~\bibnamefont
  {Bisig}}, \bibinfo {author} {\bibfnamefont {M.}~\bibnamefont {St{\"a}rk}},
  \bibinfo {author} {\bibfnamefont {M.-A.}\ \bibnamefont {Mawass}}, \bibinfo
  {author} {\bibfnamefont {C.}~\bibnamefont {Moutafis}}, \bibinfo {author}
  {\bibfnamefont {J.}~\bibnamefont {Rhensius}}, \bibinfo {author}
  {\bibfnamefont {J.}~\bibnamefont {Heidler}}, \bibinfo {author} {\bibfnamefont
  {F.}~\bibnamefont {B{\"u}ttner}}, \bibinfo {author} {\bibfnamefont
  {M.}~\bibnamefont {Noske}}, \bibinfo {author} {\bibfnamefont
  {M.}~\bibnamefont {Weigand}}, \bibinfo {author} {\bibfnamefont
  {S.}~\bibnamefont {Eisebitt}},  \emph {et~al.},\ }\href@noop {} {\bibfield
  {journal} {\bibinfo  {journal} {\textit{Nat. Commun.}}\ }\textbf {\bibinfo
  {volume} {4}} (\bibinfo {year} {2013})}\BibitemShut {NoStop}%
\bibitem [{\citenamefont {Richter}\ \emph {et~al.}(2016)\citenamefont
  {Richter}, \citenamefont {Krone}, \citenamefont {Mawass}, \citenamefont
  {Kr{\"u}ger}, \citenamefont {Weigand}, \citenamefont {Stoll}, \citenamefont
  {Sch{\"u}tz},\ and\ \citenamefont {Kl{\"a}ui}}]{Ric16}%
  \BibitemOpen
  \bibfield  {author} {\bibinfo {author} {\bibfnamefont {K.}~\bibnamefont
  {Richter}}, \bibinfo {author} {\bibfnamefont {A.}~\bibnamefont {Krone}},
  \bibinfo {author} {\bibfnamefont {M.-A.}\ \bibnamefont {Mawass}}, \bibinfo
  {author} {\bibfnamefont {B.}~\bibnamefont {Kr{\"u}ger}}, \bibinfo {author}
  {\bibfnamefont {M.}~\bibnamefont {Weigand}}, \bibinfo {author} {\bibfnamefont
  {H.}~\bibnamefont {Stoll}}, \bibinfo {author} {\bibfnamefont
  {G.}~\bibnamefont {Sch{\"u}tz}}, \ and\ \bibinfo {author} {\bibfnamefont
  {M.}~\bibnamefont {Kl{\"a}ui}},\ }\href@noop {} {\bibfield  {journal}
  {\bibinfo  {journal} {\textit{Phys. Rev. B}}\ }\textbf {\bibinfo {volume}
  {94}},\ \bibinfo {pages} {024435} (\bibinfo {year} {2016})}\BibitemShut
  {NoStop}%
\bibitem [{\citenamefont {Chiang}\ \emph {et~al.}(2010)\citenamefont {Chiang},
  \citenamefont {Chang}, \citenamefont {Yu}, \citenamefont {Huang},
  \citenamefont {Chen}, \citenamefont {Yao},\ and\ \citenamefont
  {Lee}}]{Chi10}%
  \BibitemOpen
  \bibfield  {author} {\bibinfo {author} {\bibfnamefont {T.}~\bibnamefont
  {Chiang}}, \bibinfo {author} {\bibfnamefont {L.}~\bibnamefont {Chang}},
  \bibinfo {author} {\bibfnamefont {C.}~\bibnamefont {Yu}}, \bibinfo {author}
  {\bibfnamefont {S.}~\bibnamefont {Huang}}, \bibinfo {author} {\bibfnamefont
  {D.}~\bibnamefont {Chen}}, \bibinfo {author} {\bibfnamefont {Y.}~\bibnamefont
  {Yao}}, \ and\ \bibinfo {author} {\bibfnamefont {S.}~\bibnamefont {Lee}},\
  }\href@noop {} {\bibfield  {journal} {\bibinfo  {journal} {\textit{Appl.
  Phys. Lett.}}\ }\textbf {\bibinfo {volume} {97}},\ \bibinfo {pages} {022109}
  (\bibinfo {year} {2010})}\BibitemShut {NoStop}%
\bibitem [{\citenamefont {Martinez}\ \emph {et~al.}(2007)\citenamefont
  {Martinez}, \citenamefont {Lopez-Diaz}, \citenamefont {Torres}, \citenamefont
  {Tristan},\ and\ \citenamefont {Alejos}}]{Mar07}%
  \BibitemOpen
  \bibfield  {author} {\bibinfo {author} {\bibfnamefont {E.}~\bibnamefont
  {Martinez}}, \bibinfo {author} {\bibfnamefont {L.}~\bibnamefont
  {Lopez-Diaz}}, \bibinfo {author} {\bibfnamefont {L.}~\bibnamefont {Torres}},
  \bibinfo {author} {\bibfnamefont {C.}~\bibnamefont {Tristan}}, \ and\
  \bibinfo {author} {\bibfnamefont {O.}~\bibnamefont {Alejos}},\ }\href@noop {}
  {\bibfield  {journal} {\bibinfo  {journal} {\textit{Phys. Rev. B}}\ }\textbf
  {\bibinfo {volume} {75}},\ \bibinfo {pages} {174409} (\bibinfo {year}
  {2007})}\BibitemShut {NoStop}%
\bibitem [{\citenamefont {Voto}\ \emph {et~al.}(2016)\citenamefont {Voto},
  \citenamefont {Lopez-Diaz},\ and\ \citenamefont {Torres}}]{Voto16}%
  \BibitemOpen
  \bibfield  {author} {\bibinfo {author} {\bibfnamefont {M.}~\bibnamefont
  {Voto}}, \bibinfo {author} {\bibfnamefont {L.}~\bibnamefont {Lopez-Diaz}}, \
  and\ \bibinfo {author} {\bibfnamefont {L.}~\bibnamefont {Torres}},\
  }\href@noop {} {\bibfield  {journal} {\bibinfo  {journal} {\textit{J. Phys.
  D: Appl. Phys.}}\ }\textbf {\bibinfo {volume} {49}},\ \bibinfo {pages}
  {185001} (\bibinfo {year} {2016})}\BibitemShut {NoStop}%
\bibitem [{\citenamefont {Schryer}\ and\ \citenamefont
  {Walker}(1974)}]{Walk75}%
  \BibitemOpen
  \bibfield  {author} {\bibinfo {author} {\bibfnamefont {N.~L.}\ \bibnamefont
  {Schryer}}\ and\ \bibinfo {author} {\bibfnamefont {L.~R.}\ \bibnamefont
  {Walker}},\ }\href@noop {} {\bibfield  {journal} {\bibinfo  {journal}
  {\textit{J. Appl. Phys.}}\ }\textbf {\bibinfo {volume} {45}},\ \bibinfo
  {pages} {5406} (\bibinfo {year} {1974})}\BibitemShut {NoStop}%
\bibitem [{\citenamefont {Glathe}\ and\ \citenamefont
  {Mattheis}(2012)}]{Mat12}%
  \BibitemOpen
  \bibfield  {author} {\bibinfo {author} {\bibfnamefont {S.}~\bibnamefont
  {Glathe}}\ and\ \bibinfo {author} {\bibfnamefont {R.}~\bibnamefont
  {Mattheis}},\ }\href@noop {} {\bibfield  {journal} {\bibinfo  {journal}
  {\textit{Phys. Rev. B}}\ }\textbf {\bibinfo {volume} {85}},\ \bibinfo {pages}
  {024405} (\bibinfo {year} {2012})}\BibitemShut {NoStop}%
\bibitem [{\citenamefont {Bedau}\ \emph {et~al.}(2007)\citenamefont {Bedau},
  \citenamefont {Kl{\"a}ui}, \citenamefont {R{\"u}diger}, \citenamefont {Vaz},
  \citenamefont {Bland}, \citenamefont {Faini}, \citenamefont {Vila},\ and\
  \citenamefont {Wernsdorfer}}]{Bed07}%
  \BibitemOpen
  \bibfield  {author} {\bibinfo {author} {\bibfnamefont {D.}~\bibnamefont
  {Bedau}}, \bibinfo {author} {\bibfnamefont {M.}~\bibnamefont {Kl{\"a}ui}},
  \bibinfo {author} {\bibfnamefont {U.}~\bibnamefont {R{\"u}diger}}, \bibinfo
  {author} {\bibfnamefont {C.~A.}\ \bibnamefont {Vaz}}, \bibinfo {author}
  {\bibfnamefont {J.~A.~C.}\ \bibnamefont {Bland}}, \bibinfo {author}
  {\bibfnamefont {G.}~\bibnamefont {Faini}}, \bibinfo {author} {\bibfnamefont
  {L.}~\bibnamefont {Vila}}, \ and\ \bibinfo {author} {\bibfnamefont
  {W.}~\bibnamefont {Wernsdorfer}},\ }\href@noop {} {\bibfield  {journal}
  {\bibinfo  {journal} {\textit{J. Appl. Phys.}}\ }\textbf {\bibinfo {volume}
  {101}},\ \bibinfo {pages} {09F509} (\bibinfo {year} {2007})}\BibitemShut
  {NoStop}%
\bibitem [{\citenamefont {Corte-Le{\'o}n}\ \emph {et~al.}(2015)\citenamefont
  {Corte-Le{\'o}n}, \citenamefont {Beguivin}, \citenamefont {Krzysteczko},
  \citenamefont {Schumacher}, \citenamefont {Manzin}, \citenamefont {Cowburn},
  \citenamefont {Antonov},\ and\ \citenamefont {Kazakova}}]{Cor15}%
  \BibitemOpen
  \bibfield  {author} {\bibinfo {author} {\bibfnamefont {H.}~\bibnamefont
  {Corte-Le{\'o}n}}, \bibinfo {author} {\bibfnamefont {A.}~\bibnamefont
  {Beguivin}}, \bibinfo {author} {\bibfnamefont {P.}~\bibnamefont
  {Krzysteczko}}, \bibinfo {author} {\bibfnamefont {H.~W.}\ \bibnamefont
  {Schumacher}}, \bibinfo {author} {\bibfnamefont {A.}~\bibnamefont {Manzin}},
  \bibinfo {author} {\bibfnamefont {R.~P.}\ \bibnamefont {Cowburn}}, \bibinfo
  {author} {\bibfnamefont {V.}~\bibnamefont {Antonov}}, \ and\ \bibinfo
  {author} {\bibfnamefont {O.}~\bibnamefont {Kazakova}},\ }\href@noop {}
  {\bibfield  {journal} {\bibinfo  {journal} {\textit{IEEE Trans. Magn.}}\
  }\textbf {\bibinfo {volume} {51}},\ \bibinfo {pages} {1} (\bibinfo {year}
  {2015})}\BibitemShut {NoStop}%
\bibitem [{\citenamefont {Sch\"afer}(2007)}]{Scha07}%
  \BibitemOpen
  \bibfield  {author} {\bibinfo {author} {\bibfnamefont {R.}~\bibnamefont
  {Sch\"afer}},\ }\enquote {\bibinfo {title} {Investigation of domains and
  dynamics of domain walls by the magneto-optical kerr-effect},}\ in\ \href
  {\doibase 10.1002/9780470022184.hmm310} {\emph {\bibinfo {booktitle}
  {Handbook of Magnetism and Advanced Magnetic Materials}}}\ (\bibinfo
  {publisher} {John Wiley \& Sons, Ltd},\ \bibinfo {year} {2007})\BibitemShut
  {NoStop}%
\end{thebibliography}%

\end{document}